\documentclass[aps,preprint,groupedaddress,showpacs,superscriptaddress]{revtex4}

\usepackage[dvips]{graphicx}
\usepackage{dcolumn}
\usepackage{bm}
\usepackage{epsfig}

\usepackage{color}

\begin{document}

\title{Spin injection in Silicon at zero magnetic field}

\author{L. Grenet}
\affiliation{CEA, INAC, SP2M, 17 rue des martyrs, F-38054, Grenoble, France}
\affiliation{Universit\'e Joseph Fourier, SP2M, F-38041, Grenoble France}
\author{M. Jamet}
\affiliation{CEA, INAC, SP2M, 17 rue des martyrs, F-38054, Grenoble, France}
\affiliation{Universit\'e Joseph Fourier, SP2M, F-38041, Grenoble France}
\author{P. No\'e}
\affiliation{CEA, INAC, SP2M, 17 rue des martyrs, F-38054, Grenoble, France}
\affiliation{Universit\'e Joseph Fourier, SP2M, F-38041, Grenoble France}
\author{V. Calvo}
\affiliation{CEA, INAC, SP2M, 17 rue des martyrs, F-38054, Grenoble, France}
\affiliation{Universit\'e Joseph Fourier, SP2M, F-38041, Grenoble France}
\author{J.-M. Hartmann}
\affiliation{CEA, LETI, F-38054, Grenoble, France}
\author{L. E. Nistor}
\affiliation{SPINTEC, CEA/CNRS/UJF/GINP INAC, 38054 Grenoble CEDEX}
\author{B. Rodmacq}
\affiliation{SPINTEC, CEA/CNRS/UJF/GINP INAC, 38054 Grenoble CEDEX}
\author{S. Auffret}
\affiliation{SPINTEC, CEA/CNRS/UJF/GINP INAC, 38054 Grenoble CEDEX}
\author{P. Warin} 
\affiliation{CEA, INAC, SP2M, 17 rue des martyrs, F-38054, Grenoble, France}
\affiliation{Universit\'e Joseph Fourier, SP2M, F-38041, Grenoble France}
\author{Y. Samson}
\affiliation{CEA, INAC, SP2M, 17 rue des martyrs, F-38054, Grenoble, France}
\affiliation{Universit\'e Joseph Fourier, SP2M, F-38041, Grenoble France}

\date{\today}

\begin{abstract}

In this letter, we show efficient electrical spin injection into a SiGe based \textit{p-i-n} light emitting diode from the remanent state of a perpendicularly magnetized ferromagnetic contact. Electron spin injection is carried out through an alumina tunnel barrier from a Co/Pt thin film exhibiting a strong out-of-plane anisotropy. The electrons spin polarization is then analysed through the circular polarization of emitted light. All the light polarization measurements are performed without an external applied magnetic field \textit{i.e.} in remanent magnetic states. The light polarization as a function of the magnetic field closely traces the out-of-plane magnetization of the Co/Pt injector. We could achieve a circular polarization degree of the emitted light of 3 \% at 5 K. Moreover this light polarization remains almost constant at least up to 200 K. 

\end{abstract}

\pacs{}


\maketitle

Spintronics aims at manipulating both charge and spin of carriers has been growing tremendously fast for the past ten years \cite{Wolf2001}. Indeed it has been predicted as an alternative solution to the conventional CMOS (Complementary Metal Oxide Semiconductor) technology that will reach physical limitations in a near future \cite{Zutic2004}. This subject relies on efficient spin injection into semiconductor heterostructures. Moreover for integration purposes, spin injection should be carried out electrically. So far, electrical spin injection has been widely demonstrated in III-V semiconductors using both ferromagnetic metals and magnetic semiconductors. The spin polarization of injected carriers is then given by the degree of circular polarization of the light emitted by a GaAs based spin-LED (Light Emitting Diode) \cite{Aranov1976} through dipolar selection rules. The most spectacular spin injection and optical detection were performed from BeMnZnSe \cite{Fiederling1999}, Fe \cite{Erve2004} or CoFe \cite{Jiang2005} injectors. Only recently, spin injection in silicon has been achieved using ferromagnetic metals. Spin polarized hot electrons were injected in a silicon wafer and spin detection was achieved by electrical means using two ferromagnetic electrodes in a GMR (Giant Magneto Resistance)-like geometry \cite{Appelbaum2007}. In parallel, Jonker \emph{et al.} demonstrated spin injection in the conduction band of silicon by measuring the circular polarization of a Si \textit{p-i-n} light emitting diode \cite{Jonker2007}. It should be stressed that silicon appears as the most promising candidate for spintronic circuitry and quantum computing systems since very long diffusion lengths and coherence times are predicted in this material owing to its low atomic mass (low spin-orbit coupling) and to the inversion symmetry of the crystal itself. Moreover the dominant naturally occuring isotope has no nuclear spin ruling out spin relaxation due to hyperfine interactions \cite{Tyryshkin2005,Jantsch2005}.

In this letter, we demonstrate spin injection into silicon using the remanent state of a perpendicularly magnetized ferromagnetic semi-transparent contact. The electron spin polarization is detected optically using a Si based spin-LED. The active region is a fully strained Si$_{0.7}$Ge$_{0.3}$ QW (quantum well). Compared to pure silicon spin-LEDs, the presence of germanium enhances the spin-orbit coupling which is required to observe circularly polarized light from the recombination of spin polarized injected electrons with holes. Spin injection from the Co/Pt ferromagnetic metal is carried out in the tunneling regime through an alumina barrier to overcome the conductivity mismatch obstacle \cite{Rashba2000}. Dipolar selection rules in spin-LEDs require perpendicular magnetization. In most experiments on spin injection, ferromagnetic films exhibit in-plane magnetization due to shape anisotropy. Hence strong external magnetic fields (2.2 T in the case of iron) are applied to saturate the film magnetization perpendicular to the QW plane. This may be the cause of many spurious effects such as magnetic circular dichroïsm (MCD) in the experimental set-up itself or in the spin-LED heterostructure. Moreover the Zeeman effect in the active region artificially increases spin lifetimes. Very few experiments have been carried out in remanent magnetic states (\textit{i.e.} with no applied field) into GaAs based spin-LEDs. Adelmann \emph{et al.} reported a remanent spin injection of about 5 \% at 2 K using the $\delta$-MnGa Schottky contact \cite{Adelmann2006}. Gerhardt \emph{et al.} demonstrated remanent spin injection using Fe/Tb multilayers with a circular polarization of 0.75 \% at 90 K \cite{Gerhardt2006}. Finally Sinsarp \emph{et al.} used a FePt/MgO spin injector and obtained a lower estimate of spin injection of 1.5 \% at room temperature \cite{Sinsarp2007}.
In this work, the Co/Pt ferromagnetic film exhibits strong perpendicular magnetic anisotropy allowing us to make spin injection in different remanent states.

\begin{figure}[hbt!!]
\begin{center}
\begin{tabular}{cc}
\includegraphics[angle=0.0,width=12 cm]{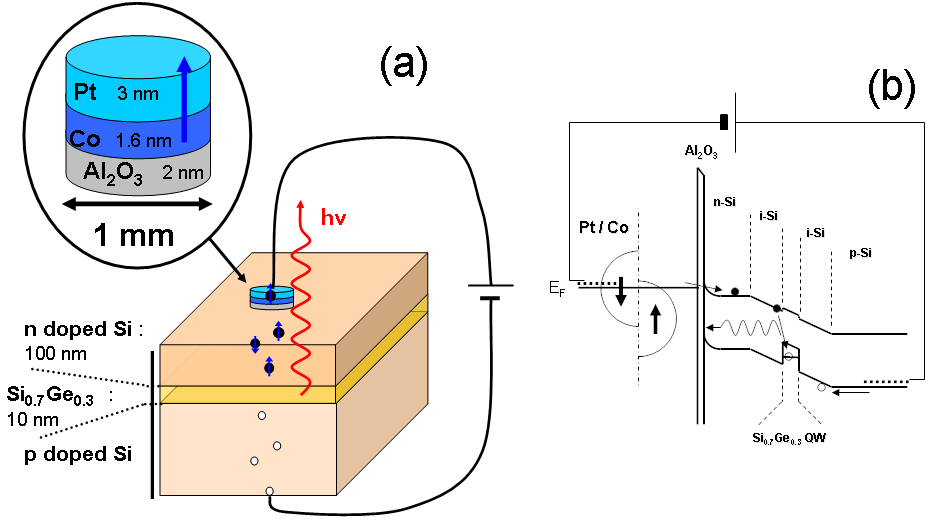}
\end{tabular}
\caption{(a) Schematic device structure of the SiGe spin-LED with the Co/Pt ferromagnetic top electrode and (b) associated simplified band diagram under bias in the injection regime.\label{spin-LED}}
\end{center}
\end{figure}
\begin{figure}[hbt!!]
\begin{center}
\begin{tabular}{cc}
\mbox{\includegraphics*[angle=0,width=12 cm]{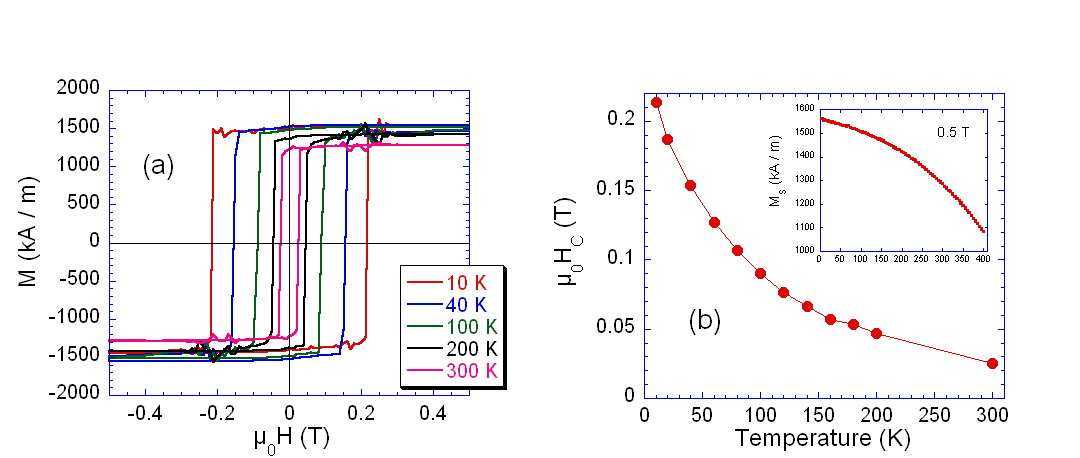}}&
\end{tabular}
\caption{(a) Hysteresis loops of the Co/Pt ferromagnetic contact recorded at different temperatures. (b) Temperature dependence of the coercive field. Inset: temperature dependence of the saturation magnetization under 0.5 T.\label{magnetism}}
\end{center}
\end{figure}

We have used Reduced Pressure - Chemical Vapor Deposition (RP-CVD) in order to grow at 26.6 mbar the following stack on a slightly p-type doped Si(001) substrate (resistivity in the 7-10 $\Omega$.cm range) : boron-doped Si ($10^{19}$ cm$^{-3}$) 500 nm / intrinsic Si (50 nm) / SiGe 30 \% (10 nm) / intrinsic Si (50 nm) / phosphorous-doped Si ($10^{18}$ cm$^{-3}$) 50 nm. The growth temperature was around 650$^{\circ}$C, in order to avoid any significant surface roughening of the Si$_{0.7}$Ge$_{0.3}$ layer during its capping with Si. The gaseous precursors used were silane (for the Si layers) and dichlorosilane + diluted germane (for the SiGe layer). For the boron (resp. phosphorous) doping, we have used diborane (resp. phosphine). High purity hydrogen with a flow of a few tens of standard litres per minute was used as the carrier gas. This SiGe based spin-LED is subsequently protected with a 2 nm-thick thermal SiO$_{2}$ layer for air transfer. Prior to the deposition of the tunnel barrier and ferromagnetic layer, this protective layer is etched using hydrofluoric acid and rinsed in deionized water. A 2 nm thick alumina tunnel barrier is then deposited at room temperature using radio-frequency (RF)-magnetron sputtering. The ferromagnetic layer is grown by direct current (DC)-magnetron sputtering at room temperature through a mechanical mask defining 900 $\mu$m wide mesas. It consists of 1.6 nm of Co covered with 3 nm of Pt. The as-grown ferromagnetic layer exhibits an in-plane magnetization. After annealing the whole heterostructure at 420$^{\circ}$C for 1 hour and 30 minutes the Co layer magnetization points out-of-plane. The whole structure as well as the band structure \cite{Weber1989} is sketched in Fig.~\ref{spin-LED}.\\
SQUID (Superconducting QUantum Interference Device) measurements with the applied field perpendicular to the film plane are displayed in Fig.~\ref{magnetism}. In Fig.~\ref{magnetism}a are shown hysteresis loops recorded at different temperatures. As shown in Fig.~\ref{magnetism}b, the coercive field strongly decreases with temperature by 88.3 \%. However low coercive fields at room temperature are required to switch easily up and down the magnetization of the spin injector in future spintronic applications. In the inset of Fig.~\ref{magnetism}b, we have plotted the temperature dependence of the saturation magnetization. The saturation magnetization is the one of bulk cobalt at low temperature. It however decreases by 30.5 \% when temperature increases from 5 K to 400 K. This temperature dependence follows a Bloch law and we could roughly estimate a Curie temperature of 900 K which is much lower than bulk cobalt (T$_{C}$=1400 K). This may be due to dimensionality effects or more probably to Co-Pt intermixing during sample annealing. Since the coercive field is correlated to the magnetization, its sharp decrease with temperature can be explained by the magnetization temperature evolution. However, we could not find a simple power law for $H_{C}$ versus $M_{S}$.

\begin{figure}[hbt!!]
\begin{center}
\begin{tabular}{cc}
\mbox{\includegraphics*[angle=0,width=0.8\textwidth]{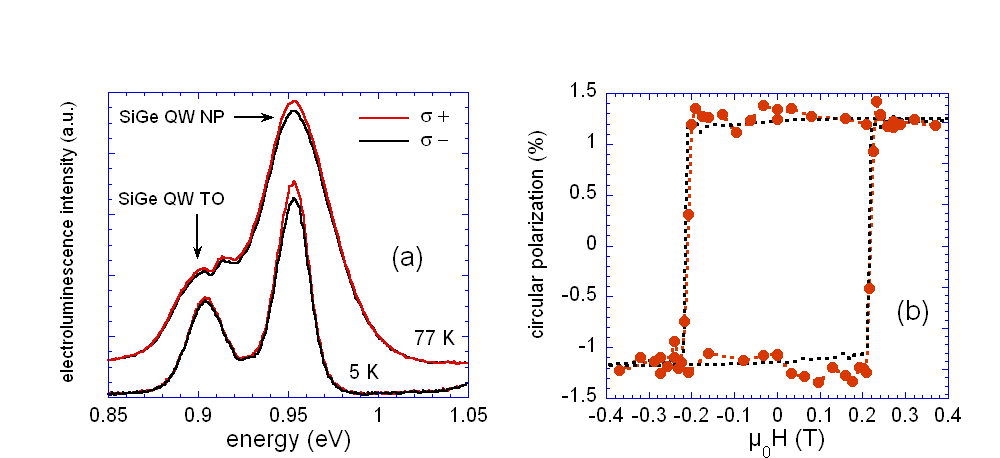}}&
\end{tabular}
\caption{(a) Electroluminescence (EL) spectra recorded at 5 K and 77 K for a saturated remanent state of the ferromagnetic Co/Pt contact. The applied voltage and current are respectively around 10 V and 10 mA. The no-phonon peak is clearly circularly polarized indicating an efficient spin injection into the silicon top layer. The TO replica is also slightly circularly polarized. (b) EL circular polarization  recorded at the maximum of the NP line for different remanent states of the Co/Pt layer (red dots). The black dotted line corresponds to the SQUID measurement performed at the same temperature.\label{electrolum}}
\end{center}
\end{figure}

We performed both photoluminescence (not shown) and electroluminescence (EL) measurements on the SiGe spin-LED at various temperatures using a cooled InGaAs photomultiplier tube (PM). The EL was excited with square electrical pulses at low frequency (63.7 Hz) with a 1:1 duty cycle and the signal was recorded using a lock-in technique. The circular polarization $P=(I_{\sigma+}-I_{\sigma-})/(I_{\sigma+}+I_{\sigma-})$ is determined using a rotatable quarter waveplate and a fixed linear polarizer placed in front of the detector. In Fig.~\ref{electrolum}a are reported EL spectra recorded at 5 K and 77 K for a saturated remanent state of the ferromagnetic contact. In addition to the silicon EL (not shown), the spectra are dominated by two main features at $\sim$950 meV corresponding to the NP (no phonon) recombination line and $\sim$900 meV corresponding to its TO phonon replica \cite{Weber1989}. Both lines are clearly circularly polarized as a consequence of momentum transfer between the spin polarized injected electrons and the emitted photons. In the following, we only consider the highest circular polarization of the NP line. We could reach a maximum of $P$=3 \% at 5 K. Moreover this circular polarization remains almost constant up to 200 K. Above this temperature, the EL signal from the SiGe quantum well becomes too weak to measure the circular polarization. In order to rule out any extrinsic effects, we performed EL measurements using a gold semi-transparent contact and inserting in the optical path a thin Si plate on which we deposited the Co/Pt ferromagnetic layer. Hence we could confirm that magnetic circular dichroism in the Co/Pt layer in a saturated remanent state is negligible. Moreover, Zeeman effect in the Si$_{0.7}$Ge$_{0.3}$ quantum well could be ruled out since the stray field from the saturated Co/Pt contact is negligible as pointed out by Gerhardt et al. \cite{Gerhardt2006}. 


In Fig.~\ref{electrolum}b, we have plotted the EL circular polarization for different magnetic remanent states of the ferromagnetic contact. In this experiment, EL measurements are performed without applied magnetic field. The reported magnetic field in the graph corresponds to the magnetic field required to reach a given remanent state of the magnetization 
of the Co/Pt layer. The circular polarization exactly traces the SQUID measurements indicating that the spin orientation of the injected electrons that radiatively recombine in the SiGe quantum well reflects that of the electron spin orientation in the Co/Pt film. 
The measured circular polarization corresponds to a lower bound of the actual spin injection in silicon. Indeed spin relaxation (with a characteristic time $\tau_{sf}$) may take place during the transit time of electrons in the silicon top layer as well as during the radiative lifetime ($\tau_{r}$) in the SiGe quantum well. Neglecting spin relaxation in the silicon top layer, the actual electron spin polarization achieved in silicon is given by $P(1+\tau_{r}/\tau_{sf})$ where the ratio $\tau_{r}/\tau_{sf}$ strongly depends on temperature and the exact electronic structure of the quantum well. We estimate the total lifetime in the quantum well : it is tens of ns, which is very short. However we have no clear estimate of the spin relaxation time ($\tau_{sf}$) in the SiGe quantum well. The determination of $\tau_{sf}$ requires time-resolved spin dynamics measurements. We can only mention electron spin resonance measurements performed by Jantsch et al. \cite{Jantsch2005} on Si$_{1-x}$Ge$_{x}$ (x$<$0.1) quantum wells grown pseudomorphically between Si$_{1-y}$Ge$_{y}$ (y$>$0.2) barriers giving $\tau_{sf}\approx$1 $\mu$s. In our case, with higher Ge content and strain, we believe that spin relaxation time should be shorter although we cannot give a lower bound. Further experiments are thus needed to estimate the actual spin injection efficiency into silicon. \\
In summary, we have shown tunneling spin injection into silicon from the remanent state of a ferromagnetic Co/Pt contact. A maximum circular polarization of 3 \% could be achieved at 5 K and remains almost constant up to 200 K. Moreover spin injection was performed for different magnetic remanent states of the ferromagnetic layer and the associated circular polarization exactly traces the SQUID measurements as expected for spin polarized electrons injected from the Co/Pt contact. 

The authors would like to thank L. Notin, A. Brenac and D. Leroy for their technical help in the sample growth.

\newpage

\end{document}